\documentclass[5p,,preprint,12pt,twocolumn]{elsarticle}
\makeatletter\if@twocolumn\PassOptionsToPackage{switch}{lineno}\else\fi\makeatother

\usepackage{ulem}
\usepackage{tabulary,xcolor}
\usepackage{amsfonts,amsmath,amssymb}
\usepackage[T1]{fontenc}
\makeatletter
\let\save@ps@pprintTitle\ps@pprintTitle
\def\ps@pprintTitle{\save@ps@pprintTitle\gdef\@oddfoot{\footnotesize\itshape \null\hfill\today}}
\def\hlinewd#1{%
  \noalign{\ifnum0=`}\fi\hrule \@height #1%
  \futurelet\reserved@a\@xhline}

\AtBeginDocument{\ifNAT@numbers \biboptions{sort&compress}\fi}
\makeatother

\usepackage{amsmath}
\usepackage{graphicx}
\usepackage[colorinlistoftodos]{todonotes}
\usepackage[colorlinks=true, allcolors=black]{hyperref}
\usepackage{float}
\usepackage{lipsum}
\usepackage{subcaption}
\usepackage{relsize}
\usepackage{dblfloatfix} 
\usepackage{kantlipsum}
\usepackage{nccmath}

\usepackage{ifluatex}
\ifluatex
\usepackage{fontspec}
\defaultfontfeatures{Ligatures=TeX}
\usepackage[]{unicode-math}
\unimathsetup{math-style=TeX}
\else 
\usepackage[utf8]{inputenc}
\fi 
\ifluatex\else\usepackage{stmaryrd}\fi

\usepackage{url,multirow,morefloats,floatflt,cancel,tfrupee}
\makeatletter

\AtBeginDocument{\@ifpackageloaded{textcomp}{}{\usepackage{textcomp}}}
\makeatother
\usepackage{colortbl}
\usepackage{xcolor}
\usepackage{pifont}
\usepackage[nointegrals]{wasysym}
\makeatletter

\def\mcWidth#1{\csname TY@F#1\endcsname+\tabcolsep}

\def\cAlignHack{\rightskip\@flushglue\leftskip\@flushglue\parindent\z@\parfillskip\z@skip}
\def\rAlignHack{\rightskip\z@skip\leftskip\@flushglue \parindent\z@\parfillskip\z@skip}

\@ifundefined{etal}{}{}

\usepackage{ifxetex}
\ifxetex\else\if@twocolumn\@ifpackageloaded{stfloats}{}{\usepackage{dblfloatfix}}\fi\fi

\AtBeginDocument{
\expandafter\ifx\csname eqalign\endcsname\relax
\def\eqalign#1{\null\vcenter{\def\\{\cr}\openup\jot\m@th
  \ialign{\strut$\displaystyle{##}$\hfil&$\displaystyle{{}##}$\hfil
      \crcr#1\crcr}}\,}
\fi
}

\AtBeginDocument{%
  \@ifpackageloaded{endfloat}%
   {\renewcommand\efloat@iwrite[1]{\immediate\expandafter\protected@write\csname efloat@post#1\endcsname{}}}{\newif\ifefloat@tables}%
}%

\def\BreakURLText#1{\@tfor\brk@tempa:=#1\do{\brk@tempa\hskip0pt}}
\let\lt=<
\let\gt=>
\def\processVert{\ifmmode|\else\textbar\fi}

\@ifundefined{subparagraph}{
\def\subparagraph{\@startsection{paragraph}{5}{2\parindent}{0ex plus 0.1ex minus 0.1ex}%
{0ex}{\normalfont\small\itshape}}%
}{}

\newcommand\role[1]{\unskip}
\newcommand\aucollab[1]{\unskip}
  
\@ifundefined{tsGraphicsScaleX}{\gdef\tsGraphicsScaleX{1}}{}
\@ifundefined{tsGraphicsScaleY}{\gdef\tsGraphicsScaleY{.9}}{}
\def\checkGraphicsWidth{\ifdim\Gin@nat@width>\linewidth
	\tsGraphicsScaleX\linewidth\else\Gin@nat@width\fi}

\def\checkGraphicsHeight{\ifdim\Gin@nat@height>.9\textheight
	\tsGraphicsScaleY\textheight\else\Gin@nat@height\fi}

\def\fixFloatSize#1{}
\let\ts@includegraphics\includegraphics

\def\inlinegraphic[#1]#2{{\edef\@tempa{#1}\edef\baseline@shift{\ifx\@tempa\@empty0\else#1\fi}\edef\tempZ{\the\numexpr(\numexpr(\baseline@shift*\f@size/100))}\protect\raisebox{\tempZ pt}{\ts@includegraphics{#2}}}}

\AtBeginDocument{\def\includegraphics{\@ifnextchar[{\ts@includegraphics}{\ts@includegraphics[width=\checkGraphicsWidth,height=\checkGraphicsHeight,keepaspectratio]}}}

\DeclareMathAlphabet{\mathpzc}{OT1}{pzc}{m}{it}

\def\URL#1#2{\@ifundefined{href}{#2}{\href{#1}{#2}}}

\def\UrlOrds{\do\*\do\-\do\~\do\'\do\"\do\-}%
\g@addto@macro{\UrlBreaks}{\UrlOrds}

\edef\fntEncoding{\f@encoding}

\makeatother

\newif\ifmultipleabstract\multipleabstractfalse%
\begin{document}
\begin{frontmatter}

\title{A geometrical model to describe the alpha dose rates \\ from particulates of UO$_2$ in water}
    
\author[a99b9588a3165]{Angus Siberry}
\ead{as14659@bristol.ac.uk}
\author[a13f753534016]{David Hambley}

\author[af593676aeb90]{Anna M Adamska}

\author[a99b9588a3165]{Ross Springell}

\address[a99b9588a3165]{
    University of Bristol\unskip, HH Wills Physics Laboratory\unskip, Bristol\unskip, BS8 1TL\unskip, UK.}
  	
\address[a13f753534016]{
    National Nuclear Laboratory Ltd\unskip, Central Laboratory\unskip, Sellafield\unskip, CA20 1PG\unskip, Cumbria\unskip, UK.}
  	
\address[af593676aeb90]{
    Sellafield Ltd\unskip, Sir Christopher Harding House\unskip, Whitehaven\unskip, CA28 7XY\unskip, UK.}
  
\begin{abstract}
A model investigating the role of geometry on the alpha dose rate of spent nuclear fuel has been developed. This novel approach utilises a new piecewise function to describe the probability of alpha escape as a function of particulate radius, decay range within the material, and position from the surface. The alpha dose rates were produced for particulates of radii 1 $\mu$m to 10 mm, showing considerable changes in the 1 $\mu$m to 50 $\mu$m range. Results indicate that for decreasing particulate sizes, approaching radii equal to or less than the range of the $\alpha$-particle within the fuel, there is a significant increase in the rate of energy emitted per unit mass of fuel material. The influence of geometry is more significant for smaller radii, showing clear differences in dose rate curves below 50 $\mu$m. These considerations are essential for any future accurate prediction of the dissolution rates and hydrogen gas release, driven by the radiolytic yields of particulate spent nuclear fuel.
\end{abstract}
\end{frontmatter}
\section*{Keywords}
Alpha Radiation, Dosimetry, UO$_2$, Radiolysis, SRIM, Geometry.

\section{Introduction}
As the demand rises for low-carbon power, the energy sector must increase renewable energy output to match \cite{outlook2020}. Nuclear power is the second-largest low-carbon energy resource globally and the largest in advanced economies supplying 18 \% of supply in 2018 \cite{sadamori2020nuclear}. The use of nuclear power has reduced CO$_2$ emissions by 74 Gt, equivalent to two years of global emissions \cite{sadamori2020nuclear,2020iaea}. Unfortunately in many advanced economies, the average age of power plants is over 35 yrs old. In Europe 83 \% of nuclear power plants are over 30 yrs old, leading to the decommissioning of approximately a quarter of the current nuclear capacity by 2025 \cite{sadamori2020nuclear}. The scale of the decommissioning to come highlights the need for innovation in reducing cost, improving monitoring, and reducing uncertainties.
A particularly important aspect of nuclear energy is the management of its highly radioactive waste, such as spent nuclear fuel. \\
\indent The fuel matrix is protected by cladding. A consistent risk to the integrity of the fuel matrix is its exposure to water in the event of cladding failure. If exposed to water, spent fuel can undergo radiolytically assisted dissolution, releasing highly radioactive fission products into the local environment \cite{shoesmith2000fuel,bobrowski2017application,bright2019comparing}. This could occur in long-term storage but is not expected (e.g. IAEA safety guide for fuel storage requires storage regimes to prevent fuel degradation), however fuel exposure in a geological repository is expected eventually \cite{shoesmith2007used, Nazhen_Shoesmith_Review}. In most countries, storage of failed fuel in ponds with no additional containment is not considered best practice. That said, it is necessary to understand what would happen if the fuel is exposed to water during storage so as to design appropriate detection and mitigation responses. The ability to predict the local environment and the mechanisms at play have the potential to increase safety and could drive down the cost of storage and decommissioning for the industry.\\ 
\indent Nuclear fuels are radially fractured during operation due to expansion stresses as they come up to power. This can expose the microstructure of the pellet surface in the event of a containment breach, post-operation \cite{fracture,shoesmith2007used}. These grains are typically 10-100 $\mu$m in size \cite{grain}. After a pellet-clad breach, water can migrate in; inter-granular radiation-induced oxidative dissolution occurs, eventually dislodging grains from the pellet, surrounding them in water \cite{shoesmith2000fuel,JonssonNielsen}.\\
\indent Ionising radiation, traversing through water, causes chemical species to form via the excitation and ionisation of the surrounding water molecules \cite{laverne2004radiation}. The process, referred to as radiolysis, greatly affects the local environment and hence the dissolution kinetics of the fuel \cite{Nazhen_Shoesmith_Review,elliot2009reaction}.\\\indent Alpha radiation is prevalent in spent nuclear fuel and is present throughout its active life cycle \cite{Nazhen_Shoesmith_Review,repository_2000}. Alpha particles can cause radiolysis and nominally travel in a straight line, leaving a penumbra of chemical events in its path \cite{elliot2009reaction}.{ 
\indent At the Sellafield nuclear legacy site in Cumbria, United Kingdom, major decommissioning projects are underway. There are four plants located at the site that contain legacy ponds and silos that store much of the UK's Magnox spent fuel such as the First Generation Magnox Storage Pond (FGMSP). The reduction of risk posed by these legacy ponds is a strategic priority of the Nuclear Decommissioning Authority (NDA) \cite{Sellafield}. 

During storage, the cladding has failed on several fuel rods, resulting in uranium `sludge' deposited at the base of the pond. Sludge at the bottom of the ponds will be a mixture of Magnox, aluminium, uranium oxide particulates, and other corrosion products \cite{sludge}. Skips with Magnox fuel are being exported to a Fuel Handling Plant (FHP) awaiting dry storage in Self-Shielded Boxes (SSBs). Uranium bit bins (UBBs) from de-cladding operations are also intended for storage in SSBs. UBBs do not drain, retaining water from wet storage, and contain insitu uranic compounds in particulate `bits'. One of the principle compounds is uranium dioxide \cite{sludge}, found in fine particulates with radii ranging from 1-50 $\mu$m. To safely store fuel in SSBs an understanding of hydrogen generation, which originates from radiolysis and uranium anoxic corrosion in the liquid water, is required.}\\
\indent To understand the radiolysis occurring locally around spent fuel, the dose rate, and dose rate profile of alpha radiation from the fuel matrix must be well-understood. To do this, predictive models are required that consider fuel activity, density, and geometry, to build the dose rate and its distribution from the surface. In this paper, a mathematical model to determine the role of geometry on the alpha dosimetry in water from a UO$_2$ surface is presented. However, this method can be applied to any spherical radionuclide-containing materials and could be widely used for a multitude of nuclear-decommissioning scenarios, in particular, worst-case risk assessments of hydrogen build-up from radiolytic production and anoxic dissolution in sludges from nuclear legacy sites, such as the Hanford K Basin sludge in the United States, and the previously mentioned Sellafield Magnox sludge \cite{k-sludge,sludge,k-sludge-gas}. 
\begin{figure}[!b]
  \centering
  \includegraphics[width=\linewidth]{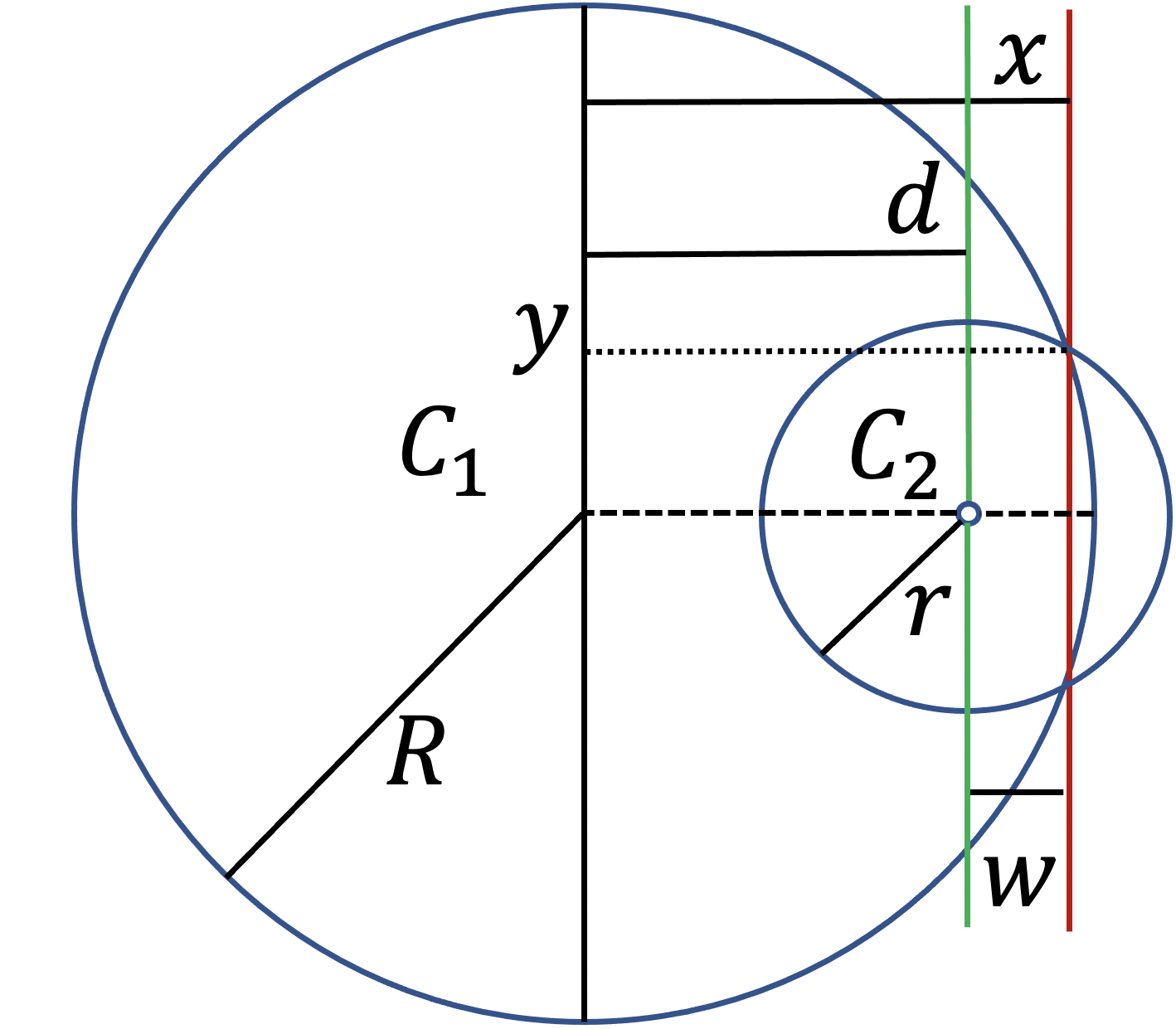}
  \caption{Diagram showing the parameters used to calculate the perpendicular distance from the origin to the radical axis (red line), $x$, of two intersecting circles $C_1$ and $C_2$ of radius $R$ and $r$, respectively. Here $d$ represents the distance between the circular centres and $w$ the perpendicular distance between the centre of circle, $C_2$, (green line) and the radical axis.}
  \label{fig:Radical_axis}
\end{figure}
\subsection{Spherical studies}
The geometrical effects on alpha dosimetry models have been researched and modelled sparingly in the last 20 years \cite{miller2006MCNP,mougnaud2015glass,poulesquen2006spherical,tribet2017spent}. The most recent attempt was made by Tribet \textit{et al.} \cite{tribet2017spent} in 2017, where spherical layers of fuel and water were used to build a Monte Carlo simulation in MCNPX. This was used to model spherical particulates of UO$_2$ with diameters 100 $\mu$m, 500 $\mu$m, and 1 mm. MCNPX is a comprehensive Monte Carlo code, which requires knowledge of the input file format, each layer to be typed in manually and is often computationally demanding. With this being the case and the significant lack of literature for small particulates of fuel in the 1-50 $\mu$m scale, a new modelling approach is required.

\section{Methods and calculations}
\subsection{Probability of alpha escape}
\begin{figure}[!b]
  \centering
  \includegraphics[width=\linewidth]{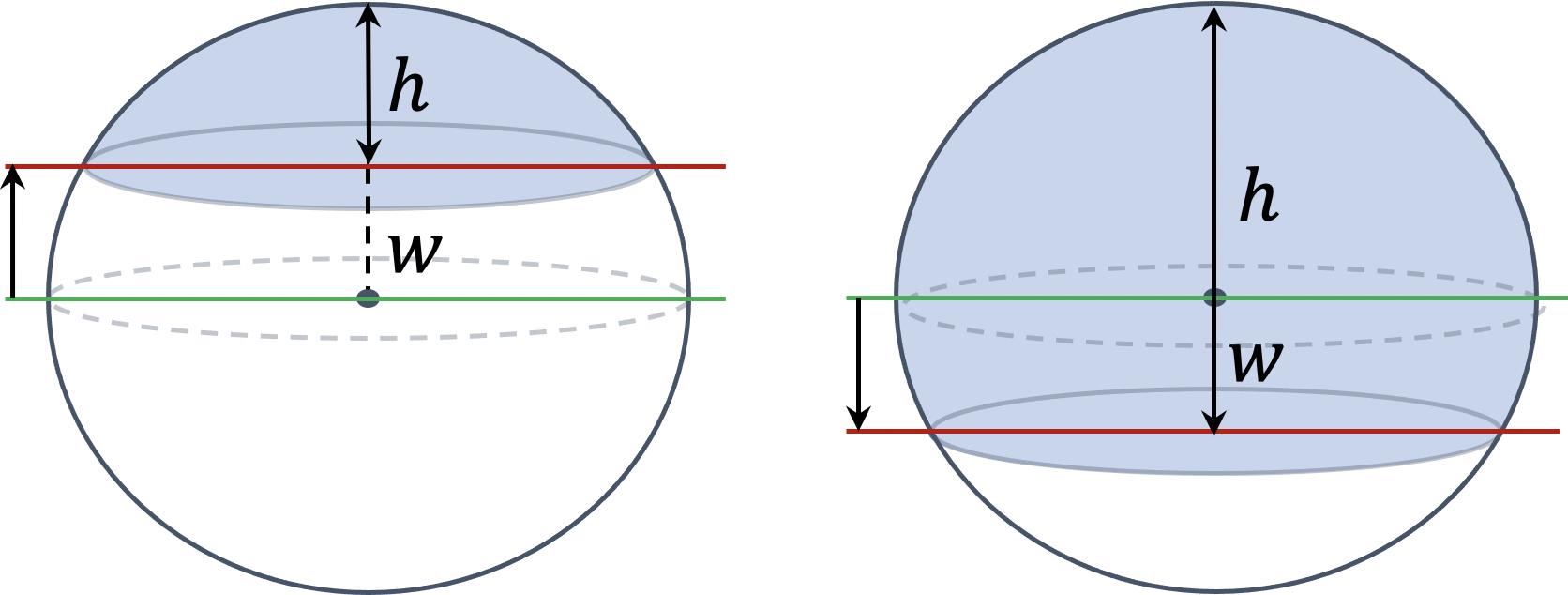}
  \caption{Diagram showing the cap height $h$ with respect to $w$. Highlighting the flip that occurs for the calculation once the radical axis moves across the centre.}
  \label{fig:h}
\end{figure}

A commonly used derivation for the probability of alpha escape for a planar surface considers a sphere and an intersecting planar surface above the sphere origin \cite{nielsen2006geometrical,dzaugis2015quantitative, SIBERRY2021109359}. The surface area ratios of the spherical shell, representing all possible decay path end positions in the fuel, and its cap beyond a planar surface represents the probability of the $\alpha$-particle passing through. To make the same surface area ratio calculation, the spherical cap size of the decay sphere must be well-understood. To calculate a spherical cap area, the intersecting planar position needs to be determined. This planar position relative to the larger sphere is known as the radical axis (see Figure \ref{fig:Radical_axis}). The radical axis represents the intersection distance of two overlapping circles from the centre of one of them. The following derivation considers two circles, the first, $C_1$, represents the fuel particulate, $C_2$ represents the possible decay end-point positions within the fuel, centred at decay distance, $d$, from the centre of the fuel particulate. Note, this also applies to spheres and includes the parameters highlighted in Figure \ref{fig:Radical_axis}.\\
\indent The equations of the two circles $C_1$ and $C_2$ centred at (0,0) and ($d$,0) respectively, are given as 
\begin{equation}
x^2 + y^2 = R^2  
\end{equation}
\begin{equation}
(x-d)^2 + y^2 = r^2.  
\end{equation}
Then substituting $y$ from both equations gives
\begin{equation}
(x-d)^2+(R^2-x^2)=r^2   
\end{equation}
where the solution for $x$ in this case is
\begin{equation}
 x = \frac{d^2-r^2+R^2}{2d}.  
\end{equation}
One can see in Figure \ref{fig:Radical_axis} that if the distance between the centre of the circle $C_1$ and the radical axis is $x$, the distance from the centre of the circle $C_2$ to the same radical axis is
\begin{equation}
  w = |x-d|.
\end{equation}
The modulus is used in this case because the radical axis distance can be closer than the distance between the two circles. It should be noted for this to be applied to the physical scenario of a $\alpha$-particle decaying from a sphere of radius R with a `decay shell' of radius $r$ the decay shell centre is bound by the solid fuel sphere. So setting the limits $r \leq R$ we can now derive the surface area of the spherical cap as a function of $d$, $S_{cap}(d)$. To do so we must define the spherical cap height as
\begin{equation}
  h = r - w
\end{equation}
or more precisely
\begin{equation}
  h(d) = r - | \frac{d^2-r^2+R^2}{2d} - d|.
\end{equation}
The surface area of a spherical cap of radius, $r$, is given as
\begin{equation}
2\pi r h  
\end{equation}
where the height is defined as the distance from, the edge of the sphere perpendicular to the plane, to the plane itself. This is shown in Figure \ref{fig:h}.

Because the surface of the material is curved the radical axis can lie closer to the centre of the material than the decay position, causing $S_{cap}(d)$ to become a piecewise function
\[
 {S_{cap} =
 \begin{cases}
        \text{$2\pi r h(d),$} & \text{$R-r < d < \sqrt{R^2 - r^2}$  } \\
\text{$4\pi r^2-2\pi r h(d),$} &\text{$\sqrt{R^2 - r^2} < d < R$.}
 \end{cases}}
\]
The limits here were derived from when the cap height exceeds the radius and the function `flips' decreasing in area past the halfway point of the sphere. This was determined by finding the inflection point of $h(d)$,where $\frac{\partial h(d)}{\partial d} = 0$, and solving for $d$.
\begin{figure*}[!b]
  \centering
  \includegraphics[width=\linewidth]{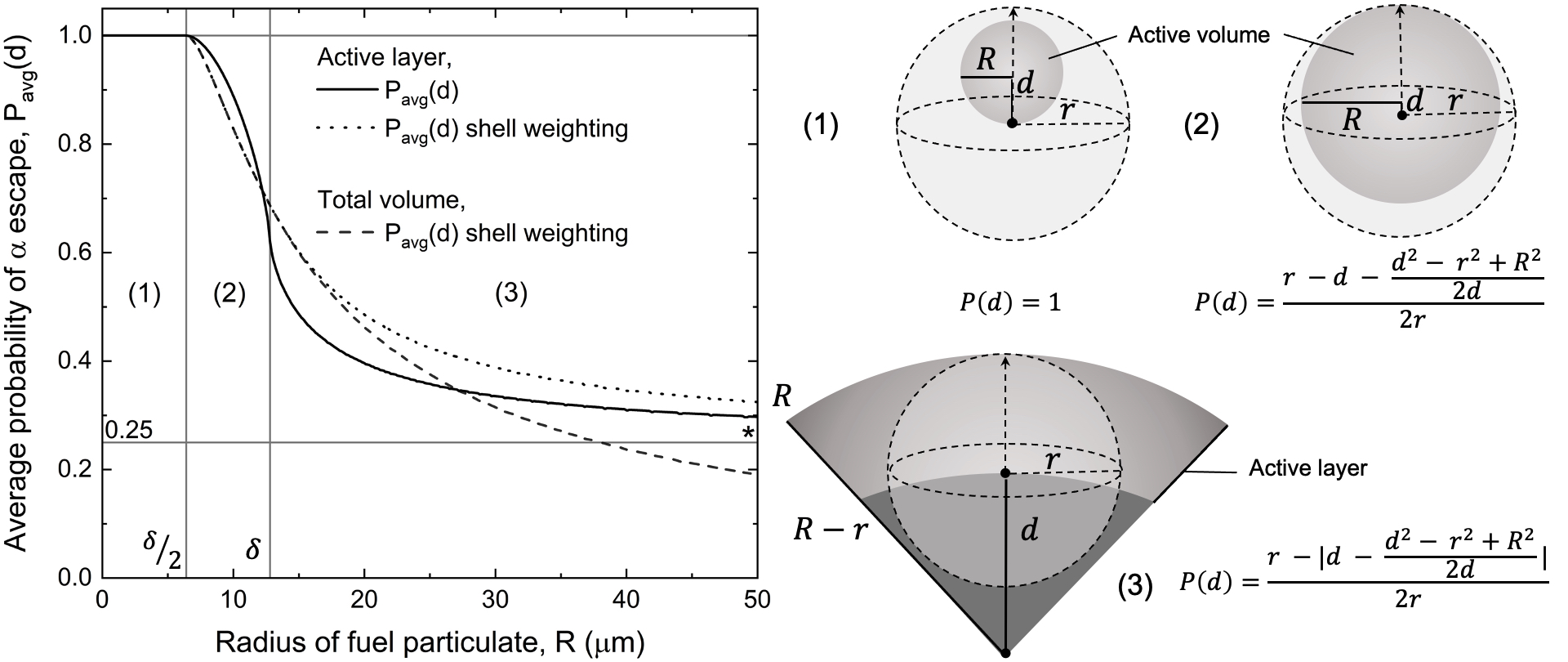}
  \caption{Graph showing the relationship between the average probability of escape and size of fuel particulate, in this case, a UO$_2$ particulate where $\delta$ = 12.8 $\mu$m. The equations 1, 2, and 3, annotated for the regime of R for which they are relevant, are used to calculate the probability of escape for a decay position from the particulate centre, $d$, within the active layer of a fuel particulate of radius, $R$. The term `shell weighting' refers to the weighting of $d^2$ to represent the shell of possible positions at distance, $d$, in 3D. The $^*$ marks the asymptotic value 0.25, representing P$_{avg}$($d$) as R $\rightarrow$ $\infty$, (an infinitely planar surface) within the active volume.}  \label{fig:Palpha}
  \vspace{2mm}
\hrule
\end{figure*}

Now we know how the surface area of the decay sphere outside of the material evolves as a function of decay position, $r$, we can determine the probability of alpha escape. Using the same surface area ratio used for a planar geometry \cite{SIBERRY2021109359}, we get the probability of alpha escape, $P(d)$, as

\begin{equation}
  P(d) = \frac{S_{cap}(d)}{4\pi r^2}.
\end{equation}

Substituting $S_{cap}(d)$ and dividing through gives
\[
 P(d) =
 \begin{cases}
        \textit{h(d)}$/2r,$ & \text{$R-r < d < \sqrt{R^2 - r^2}$} \\
1 - \textit{h(d)}$/2r,$ &\text{$\sqrt{R^2 - r^2} < d < R$.}
 \end{cases}
\]

Note that this derivation has made the assumption the particulate of fuel is larger than the `decay sphere'. When the decays sphere is larger than the particulate itself it will increase the probability of escape until the decay sphere radius, $r$, is twice the size of the radius of the particulate, 2R, converging the probability to be 1 at all possible, $d$ positions. To map the function for particulates of sizes between $r/2$ and $r$ an adjustment on the calculation for $R>r$ needs to be made. For $R \leq r$ the decay position is bound in a way that the $x$ is always negative for a positive $d$, so the function loses its modulus and becomes
\begin{equation}
  w = d + x
\end{equation}
where $x$ is the negative of (4). Following the same derivation of $h(d)$ leads to the probability distribution as a function of position, $d$, becoming
\begin{equation}
P(d) = \frac{r - d - (\frac{d^2 - r^2+R^2}{2d})}{2r}
\end{equation}
where
\[
 P(d) =
 \begin{cases}
1 - \textit{h(d)$/2r$,} &\text{$|R - r| < d < R$} \\
1. &\text{$ d < |R - r|$.}
 \end{cases}
\]
\begin{figure*}[b!]
  \centering
  \includegraphics[width=\textwidth]{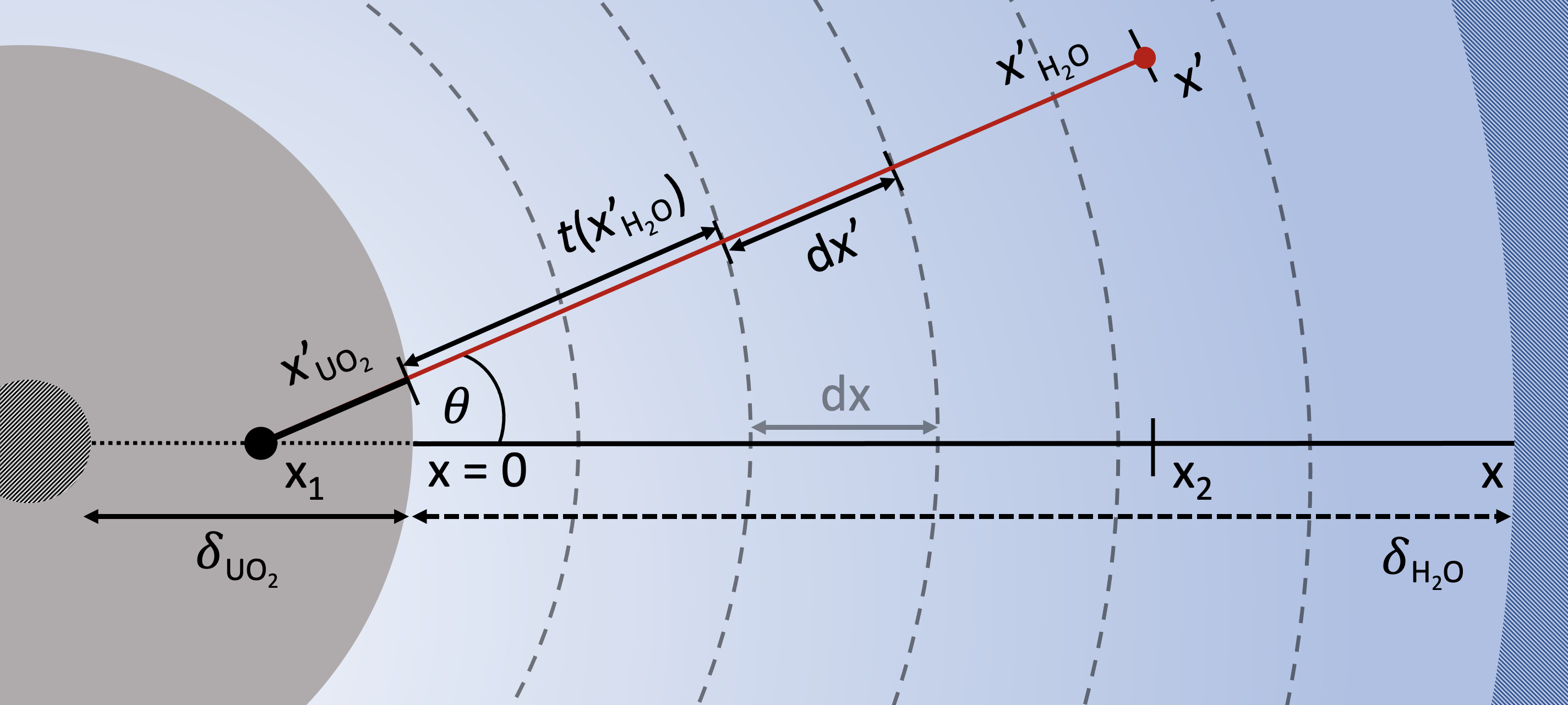}
  \caption{Illustration of the spherical geometry layer set-up, where t is the coefficient used to calculate intersections of each spherical layer. The x$_1$ position indicates a randomly generated position on the dotted line, while the x$_2$ position indicates the perpendicular range of the $\alpha$-particle. The solid black-red line denoted x'$_{UO_2}$ and x'$_{H_2O}$ indicates the distances travelled along the axis x'; a randomly generated path at an angle $\theta$ from the x-axis, in UO$_2$ and H$_2$O, respectively. dx and dx' represent infinitesimal distances between successive layers in H$_2$O.}
  \label{fig:spherical}
  \vspace{4mm}
\hrule
\end{figure*}
Therefore, the final mean probability of alpha escape is the integral of all possible $P(d)$ values within the limits of $d$, for any given $R$. The analytical result for average values of P($d$) within the active volume at a given R are shown in Figure \ref{fig:Palpha} as P(R). Furthermore, as the average value in 3D must also consider the `shell' area of positions at a given depth, the weighted average (by $d^2$) is also presented. Now considering the entire volume of the particulate, a plot can be made of P(R) as a function of particulate size. This is included in Figure \ref{fig:Palpha}, and although it is not required in the dose simulation, it is useful if one wanted to analytically calculate the  emission rate escaping a particulate of a given radius. For simplicity, the maximum decay distance, $\delta_{UO_2}$, will now be used to physically represent $r$ for the remainder of this report. The volume of material that is considered to contribute alpha decays crossing the fuel-water interface is commonly referred to as the active volume (bound by layer thickness $\delta_{UO_2}$) or active mass \cite{nielsen2006geometrical}. In this case, we can clearly see for an infinitely large circle the surface curvature with respect $\delta_{UO_2}$ tends to a planar surface with an average probability of escape of 0.25, within the active volume \cite{SIBERRY2021109359,nielsen2006geometrical,dzaugis2015quantitative}. It should also be noted that when the radius of the fuel particulate is less than half of $\delta_{UO_2}$, the alpha decay can occur at any position within the particulate and escape. It is important for the accuracy of this model to obtain the analytical solution for the probability of alpha escape. Previous Monte Carlo methods have used accept-reject methods for $\alpha$-particle paths, only using the paths that cross the fuel-water interface in the solution. This method leads to wasted information and inefficiency of computations when having to consider paths that do not cross the fuel-water interface. To generate paths that cross the fuel-water interface only, the analytical value for the probability of alpha escape is required. This method requires all $\alpha$-particle paths to escape stochastically within geometrical limits and then utilise the analytical escape probability in the  emission rate calculation. Therefore, the probability function can be considered to increase the efficiency of an accept-reject algorithm by the percentage of paths that would not escape. This function is also a useful test for the accuracy of the geometrical set-up (Figure \ref{fig:spherical}).

\subsection{Geometry}
Calculating the distance travelled within the particulate, $\delta_{UO_2}$ is the most important step in this model as it allows us to determine how much energy the $\alpha$-particle has when leaving the sphere and hence its respective dose contribution to the surrounding water. In a planar surface dosimetry model by Siberry \textit{et al.} \cite{SIBERRY2021109359}, the Stopping Ranges of Ions in Matter (SRIM) software is used to extract a Bragg curve of a chosen decay energy through water \cite{ziegler2008srim}. The model utilises the Linear Energy Transfer (LET) output as a `base function' for all $\alpha$-particle paths in water by assuming a linear shift in range with distance travelled through UO$_2$. Using the value of $x'_{UO_2}$ and the method presented by Siberry \textit{et al.}, the `base function' can be manipulated to represent the LET for this $\alpha$-particle path beyond the fuel-water interface.

\begin{figure}[!b]
  \centering
    \includegraphics[width=\linewidth]{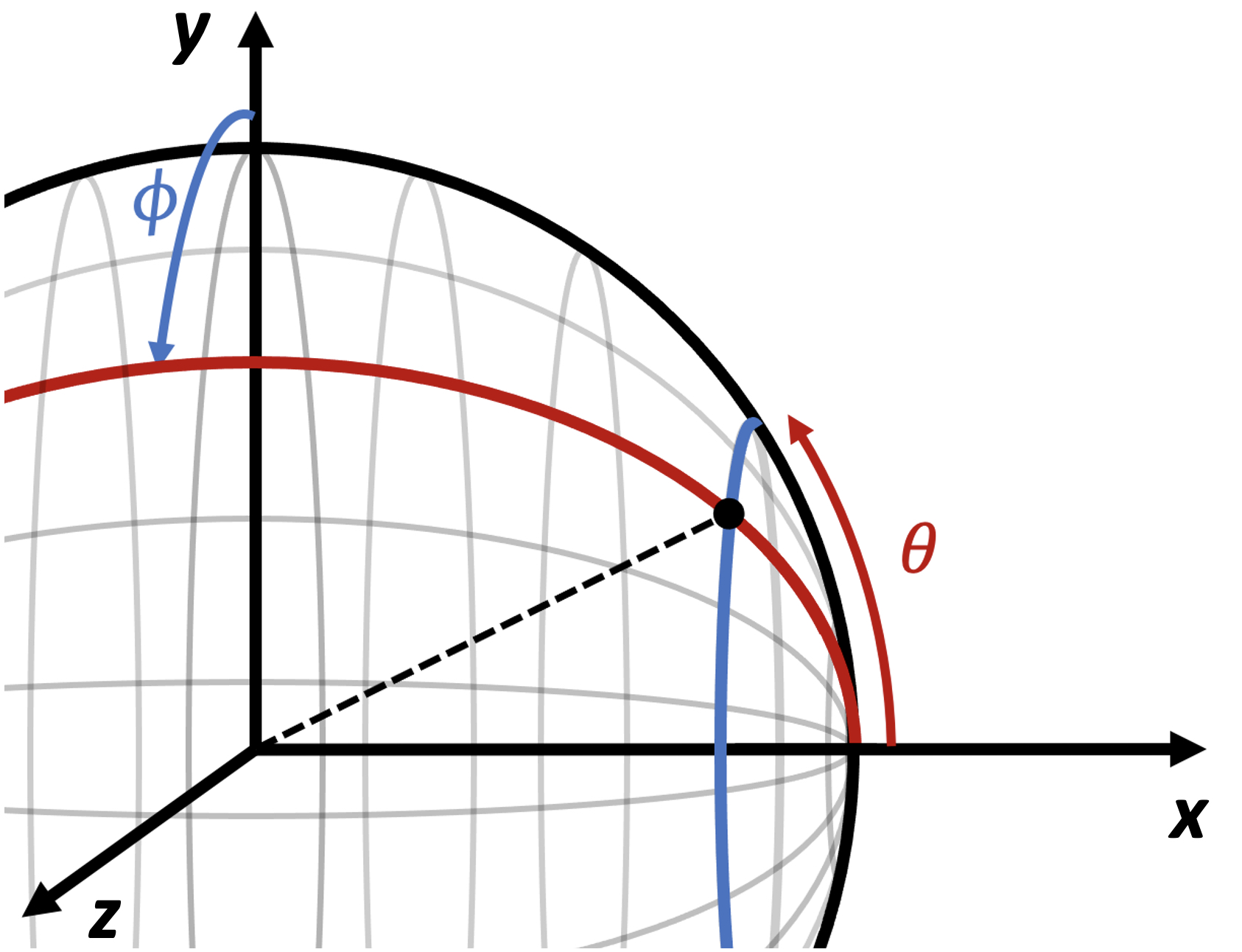}
  \caption{ Diagram illustrating the role of $\theta$ and $\phi$ on the the choice of random paths in 3D using sphere point picking methods.}
  \label{fig:3D}
\end{figure}

\begin{figure}[!t]
  \centering
  \includegraphics[width=\linewidth]{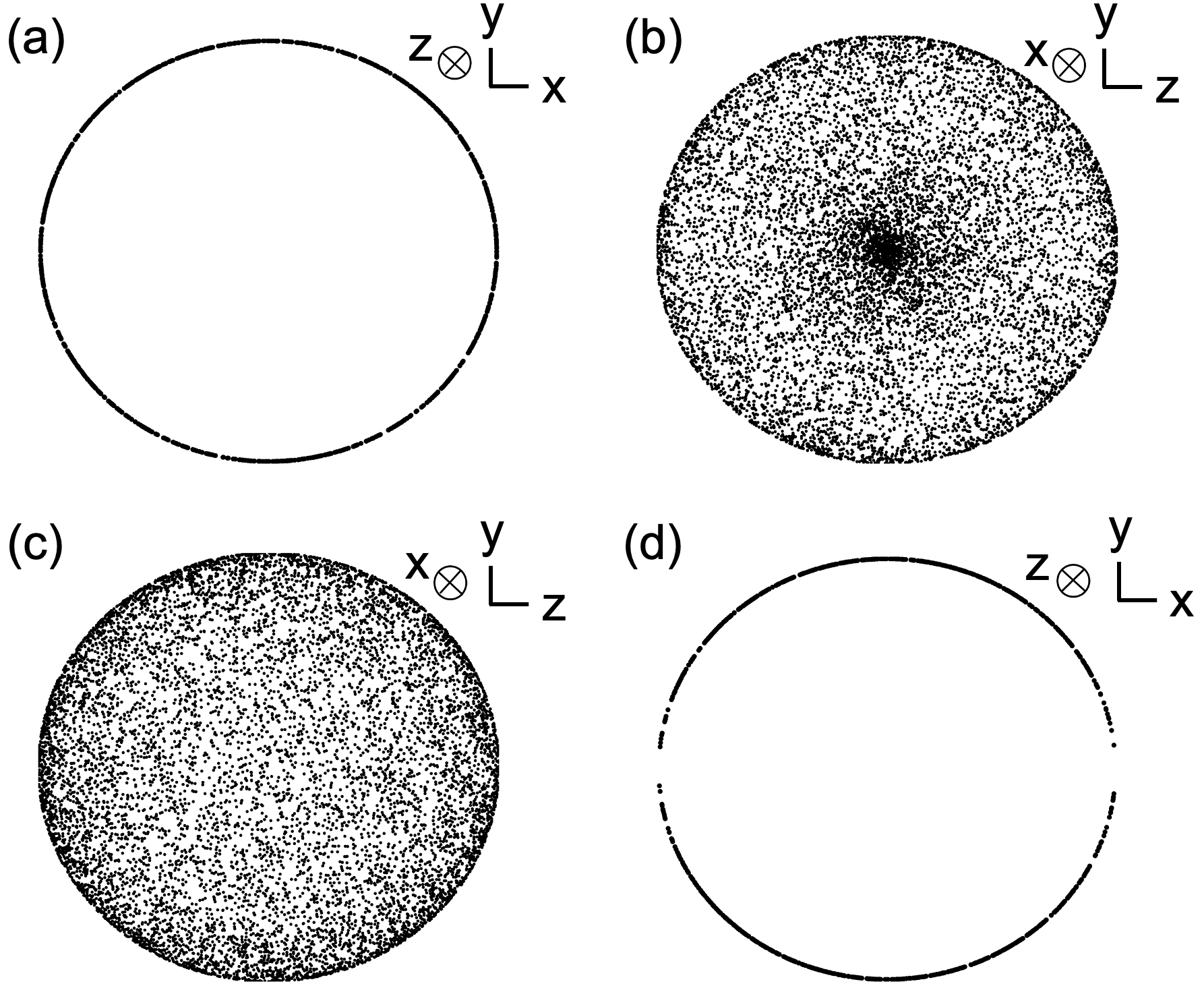}
  \caption{{Coordinate plots of randomly distributed points, using a variety of sampling methods: (a) uniformly random $\theta$ angles with $\phi = 0$, (b) uniformly random $\phi$ and $\theta$ angles in the point of view parallel to the z-y plane, (c) uniformly distributed points using a sphere point picking method proposed by Weisstein, 2002 \cite{weisstein2002sphere}, parallel to the z-y plane, (d) the same sphere point picking method, but rotating all points around the azimuthal axis until $\phi = 0$, producing a uniformly distributed sphere point picking scheme in 2D.}}
  \label{fig:point}
\end{figure}

\subsubsection{2D sphere point picking}
In order to represent a true sphere, contributions from the third dimension must be accounted for to represent all possible decay paths (Figure \ref{fig:3D}). When choosing a random angle for the decay path, intuitively one can take a uniformly random number to represent an angle $\theta$ between 0 and 2$\pi$ to generate an evenly distributed uniform circle of points, see Figure \ref{fig:point}(a). Applying the same logic for three dimensions one would use the same method for a uniformly random $\phi$ angle in the azimuthal direction. Figure \ref{fig:3D} illustrates the process for picking points on a sphere. We see that the distribution of randomly allocated points about the azimuthal direction will have a higher density of points about the pole, in this case lying along the x-axis, resulting in the non-uniform distribution shown in Figure \ref{fig:point}(b). There have been several techniques derived to uniformly distribute points on the surface of a sphere \cite{von195113,cook1957rational,marsaglia1972choosing,weisstein2002sphere}. The most elegant solution uses the cos-inverse of a uniform random number that then allows a uniform normal distribution over the azimuthal angle to be taken \cite{weisstein2002sphere}. The result of the distribution is shown in Figure \ref{fig:point}(c). Due to the axial symmetry of the geometry used in this model, simplifications can be made. In the case of path distribution, the sphere point picking algorithm is required but in the absence of a z-component. This can be achieved by taking each particle position in 3D and rotating it about the azimuthal direction until $\phi = 0$. The result of this is a sphere point picking distribution in 2D shown in Figure \ref{fig:point}(d). As expected there is a higher density of points about the y axis to counter the spherical bunching at the poles in Figure \ref{fig:point}(b), ensuring uniformity in the azimuthal direction.

\subsubsection{Dose calculation}
The model generates a random starting depth within the fuel bound by the range limit of the $\alpha$-particle of a given energy within the material, $\delta_{UO_2}$. The angle, $\theta$, is generated within limits to ensure the $\alpha$-particle escapes. Once the trajectory and escape energy are known, the model constructs layers inside the water and calculates the LET value at each interval, approximating a fixed energy transfer across each layer. In the spherical case, the layers are spherical shells of increasing radius as shown in Figure \ref{fig:spherical}. The path length across each layer varies depending on the angle of incidence from the axis of symmetry. Hence, the distance the alpha decay travels within each layer is non-uniform resulting in a new path distance to be calculated at each interval. At each $x$ coordinate, the $x'$ value on the LET curve is multiplied by a thickness $dx'$ to be deposited in the interval $x + dx$. This process is done until the alpha decay reaches its stopping range (LET = 0). At each layer, the values are deposited into an array, which provides the dose rate as a function of distance from the interface. The simulation then starts again, until the user-defined number of $\alpha$-particles are simulated. In order to produce a dose rate, the result must represent the number of alpha decays within a given time interval. To do this, the average dose per gram of water layer is produced per decay. This output can then be scaled in post-analysis to represent the required dose rate.

\begin{figure}[!b]
  \centering
  \includegraphics[width=\linewidth]{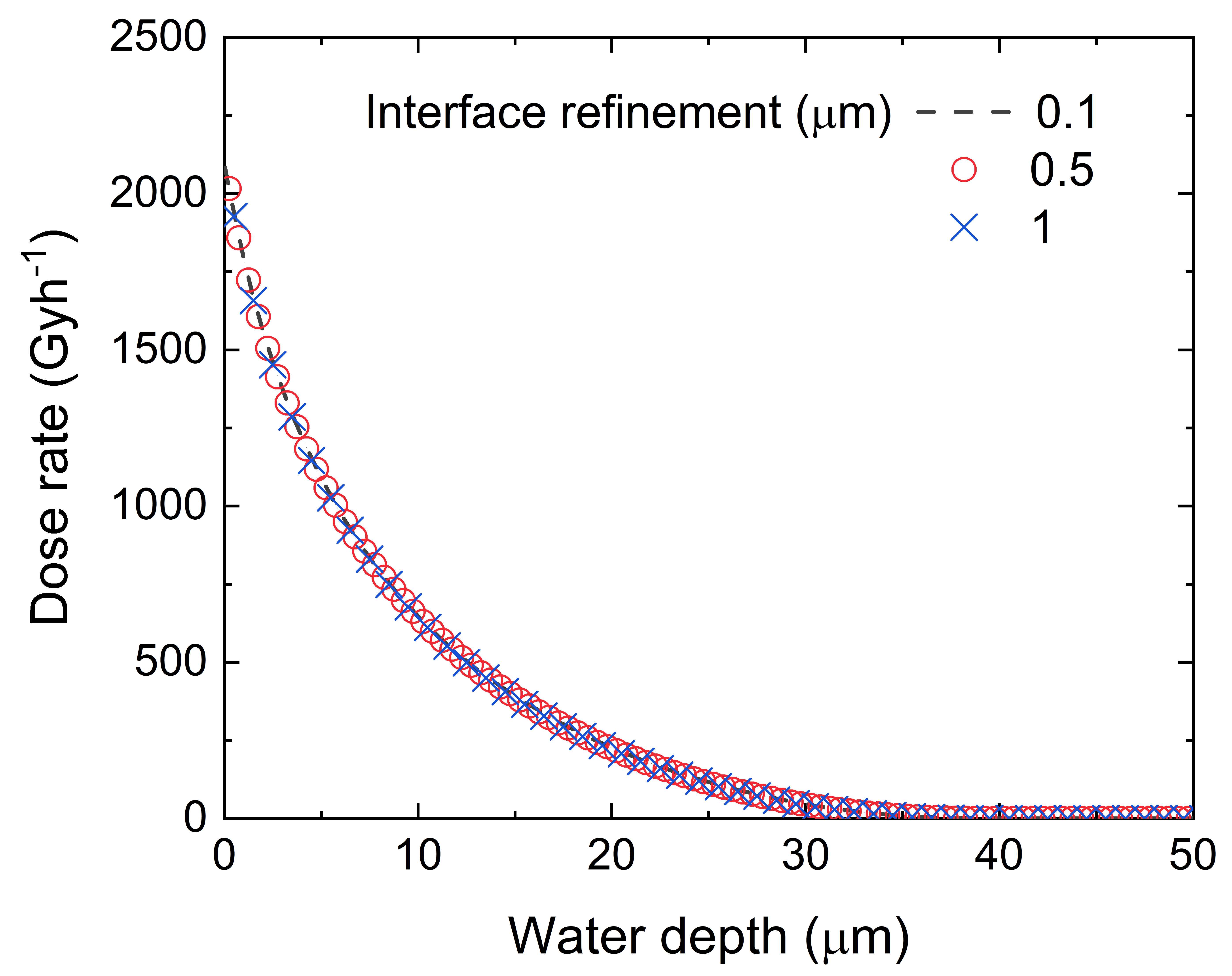}
  \caption{Comparison of dose rate profiles for 1 $\mu$m, 0.5 $\mu$m and and 0.1 $\mu$m layer step sizes for a 50 $\mu$m radius UO$_2$ sphere.}
  \label{fig:step}
\end{figure}

\section{Results}
\subsection{Simulation analysis}
To test this 2D geometrical approach over the 3D Monte-Carlo method presented by Tribet \textit{et al.}, the same source particle number (SPN), 500,000 and interface refinement steps, 0.1 $\mu$m, were used. The source code of the Tribet model uses a random distribution of points within the whole sphere, where only 16 \% of decays will cross the interface for a 50 $\mu$m radius particulate. The model presented in this study generates randomly positioned starting points (weighted by $d^2$) within the active layer, then calculates the theta limits in which each $\alpha$-particle can escape before randomly generating an angle with the weighted distribution described in Figure \ref{fig:point}.\\
The same unit conversion is used between this study and Tribet \textit{et al.},

\begin{equation}
\dot{D} = Ed_0 \cdot A \cdot C
\end{equation}

where; $\dot{D}$ represents the dose rate, $Ed_0$ represents average dose from a single decay through each layer (in MeVg$^{-1}$), $A$ is the activity of the source (in Bq) and $C$ is the conversion factor, 5.76$\times 10^{-7}$ (in h$^{-1}$sJMeV$^{-1}$gkg) \cite{tribet2017spent}. The output of this model represents the dose from the SPN. The dose rate is then divided by the SPN to describe the average LET. The LET is then converted to MeV$g^{-1}$ and the analytical P$_{\alpha}$ value is used to calculate the appropriate  emission rate value. The key difference is the use of P$_{\alpha}$ in the activity of the source as that information is not stored within the $Ed_0$ value. To represent a dose-rate curve with the same SPN of 500,000 in MCNPX, the present study needs only 81,800 randomly generated particle paths to converge to the same result for a particulate of 50 $\mu$m radius. Another factor to consider is the use of 2D discretised particle paths instead of 3D, for example, if a refinement of 0.1 $\mu$m layers was implemented, intersection point calculations would be in the order of 600 million (500,000 $\cdot$ 400 $\cdot$ 3) as opposed to 65 million (81,800 $\cdot$ 400 $\cdot$ 2), not including any other dimensional calculations required for 3D positional arrays. Suggesting that, depending on the use of programming language a considerable increase in computational efficiency. The use of a FORTRAN-based code such as MCNPX makes the user input a multitude of input file cards, and using the same method by Mougnaud \textit{et al.}, each layer must be enteredmanually. This entire model can be implemented in a multitude of languages, such as Python, C++, Java allowing for user-friendly interfaces to be developed for simple and fast calculations. This format also allows for flexibility and compatibility when implementing as a module in more complex time-resolved multi-physics or chemical kinetic models.\\
\begin{figure}[!b]
  \centering
  \includegraphics[width=\linewidth]{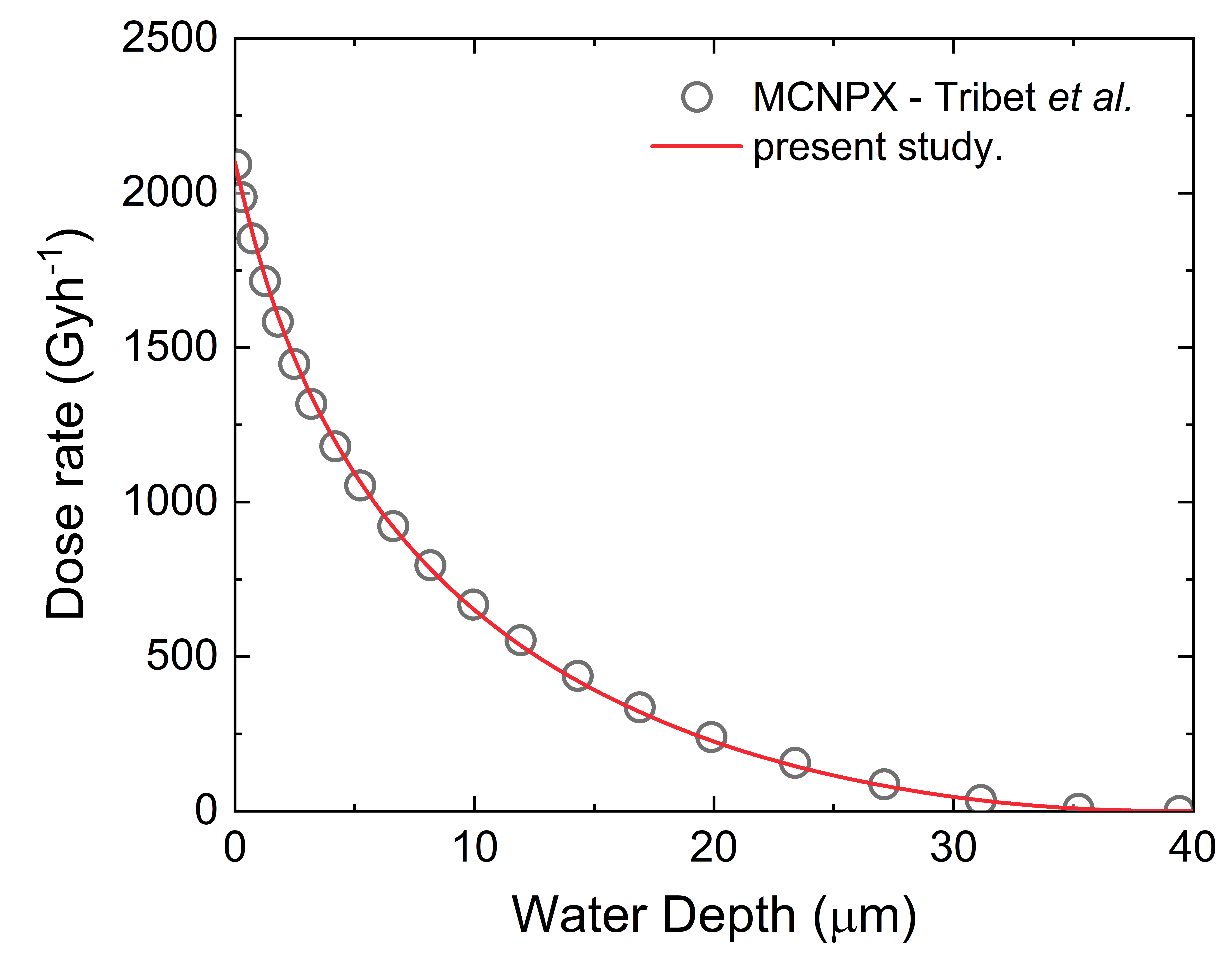}
  \caption{A comparison of the dose rate profiles derived from the present study and the MCNPX model by Tribet \textit{et al.}, for a particulate of radius 50 $\mu$m. The activity and energy used in both studies were 4.73$\times 10^8$ Bqg$^{-1}$ and 5.3 MeV, respectively.}
  \label{fig:Tribet_comp}
\end{figure}
\begin{figure}[!b]
  \centering
  \includegraphics[width=\linewidth]{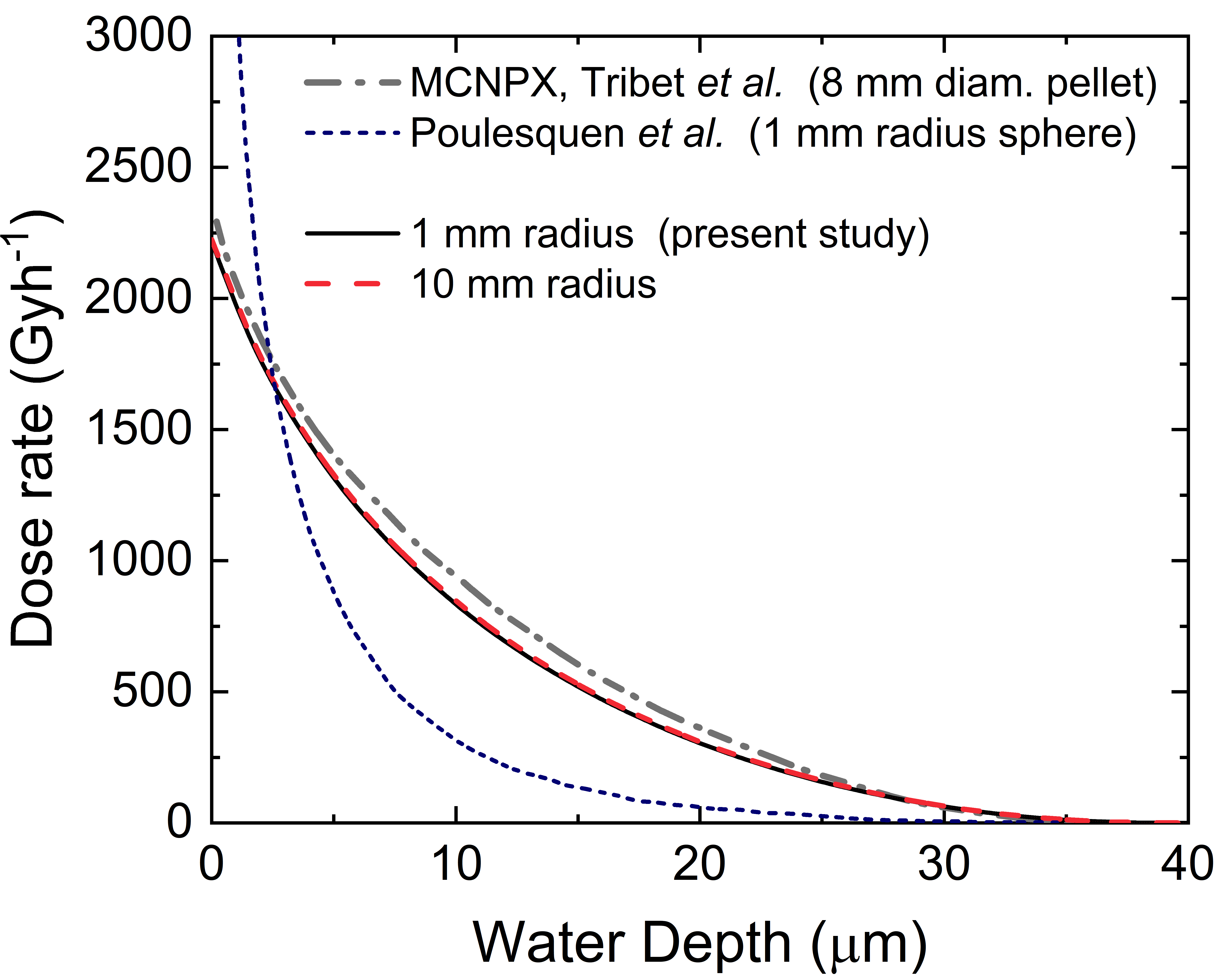}
  \caption{ Dose rate profiles for large particulates where the radius, R$>>\delta$, compared to results from Tribet \textit{et al}. and Poulesquen \textit{et al}. The activity and energy used in all studies were 4.73$\times 10^8$ Bqg$^{-1}$ and 5.3 MeV, respectively.}
  \label{fig:large_comp}
\end{figure}
\begin{figure*}
  \centering
  \includegraphics[width=\linewidth]{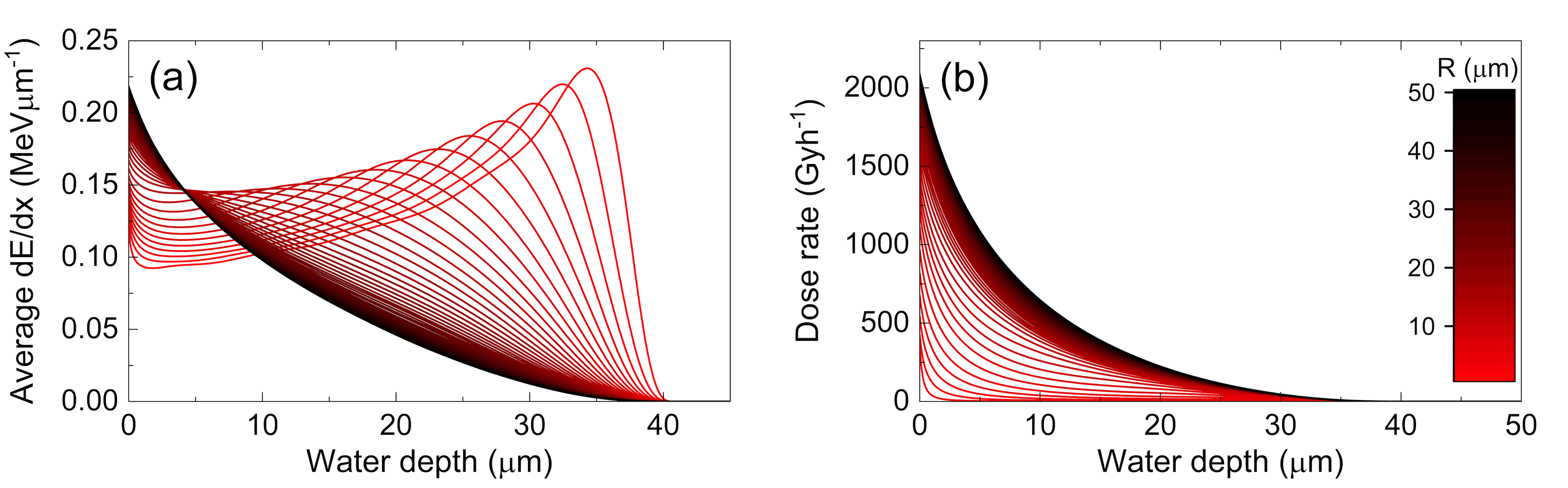}
  \caption{Results for particulates ranging from 1-50 $\mu$m in radius where (a) is the average energy deposited in the surrounding spherical layers per $\alpha$-decay and (b) is the dose rate profile per particulate. The activity and energy used in all simulations were 4.73$\times 10^8$ Bqg$^{-1}$ and 5.3 MeV, respectively.}
 \label{fig:full_results}
\vspace{4mm}
\hrule
\end{figure*}
\noindent The effect of interface refinement step value on the dose rate curve can be seen in Figure \ref{fig:step}. Varying interface refinement steps by 1 $\mu$m, 0.5 $\mu$m and 0.1 $\mu$m show no significant modification in resulting dose rate curves.

\subsection{Model assumptions and limitations}
The limitations of this model are that: 1) it relies on SRIM output data accurately representing the decay path through both UO$_2$ and liquid water; and 2) the path characteristics are uniform in nature, assuming that $\alpha$-emitters are distributed homogeneously in the medium, that emissions are isotropic, and neglecting scattering events that change the direction of the $\alpha$-particle velocities.

\subsection{Comparison with previous literature}
The MCNPX model by Tribet \textit{et al.} produces the only published dose profile for particulates with R < 100 $\mu$m, therefore a comparison must be made between the two models utilising the same input parameters. The results of the Tribet model and present study for a UO$_2$ particulate of radius, R = 50 $\mu$m and oxide density of 10.8 gcm$^{-3}$ are shown in Figure \ref{fig:Tribet_comp}. The resulting dose rate curves show a very strong correlation in decay shape, range, and interface dose rate. The total dose rate over the 40 $\mu$m thick water layer receiving the dose is within 0.7 \% of the result by Tribet.\\
\indent Dose rate profiles were produced for particulates of a radius of curvature much greater than the decay sphere radius, $\delta_{UO_2}$ (Figure \ref{fig:large_comp}). The dose rates produced were calculated using an activity of 4.73$\times 10^8$ Bqg$^{-1}$ and decay energy of 5.3 MeV to compare with models produced by Poulesquen and Tribet \cite{poulesquen2006spherical,tribet2017spent}. The model converges at larger radii, as expected, and the decay shape agrees closely with the result produced by Tribet \cite{tribet2017spent}.\\

\subsection{Role of particulate size}
The effect of particulate size in this study was examined (see Figure \ref{fig:full_results}). To better understand the energy transfer to the local environment, the average LET was plotted for each particulate size (Figure \ref{fig:full_results}a). The average LET for the 1 $\mu$m result represents a typical Bragg curve of an $\alpha$-particle in water due to the short distances travelled within UO$_2$ (< 1 $\mu$m). In this case, most of the dose is deposited at the end of its path. This functional form then decomposes with larger radii, better representing a decay shape towards R = 50 $\mu$m. Figure \ref{fig:full_results}(b) shows the dose rate across the surrounding water layers from each particulate. The curves differ from Figure \ref{fig:full_results}(a) to Figure \ref{fig:full_results}(b) because the LET function is suppressed by successive spherical layers of increasing volume.\\
\begin{figure}[!b]
\centering
\includegraphics[width=\linewidth]{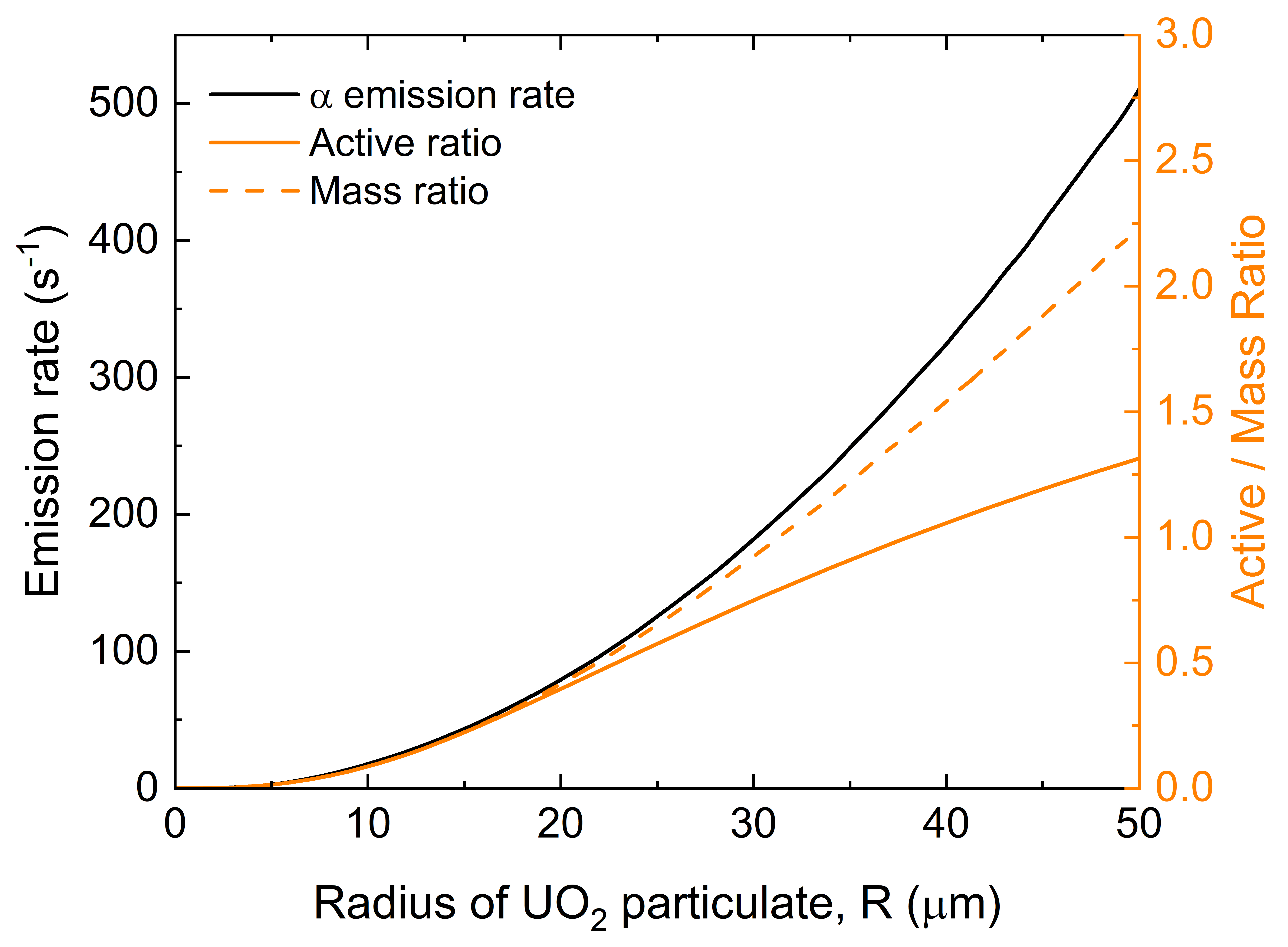}
 \caption{Graph illustrating the role of UO$_2$ particulate size on dosimetry conditions. The black line indicates the rate of alpha particles emitted per particulate. The orange lines indicate the ratio of mass from the active layer (full line) and entire particulate (dashed line) mass to the mass of affected water volume, denoted Active ratio and Mass ratio, respectively.}
 \label{fig:perp}
\end{figure}
\begin{figure}[!b]
\centering
 \includegraphics[width=\linewidth]{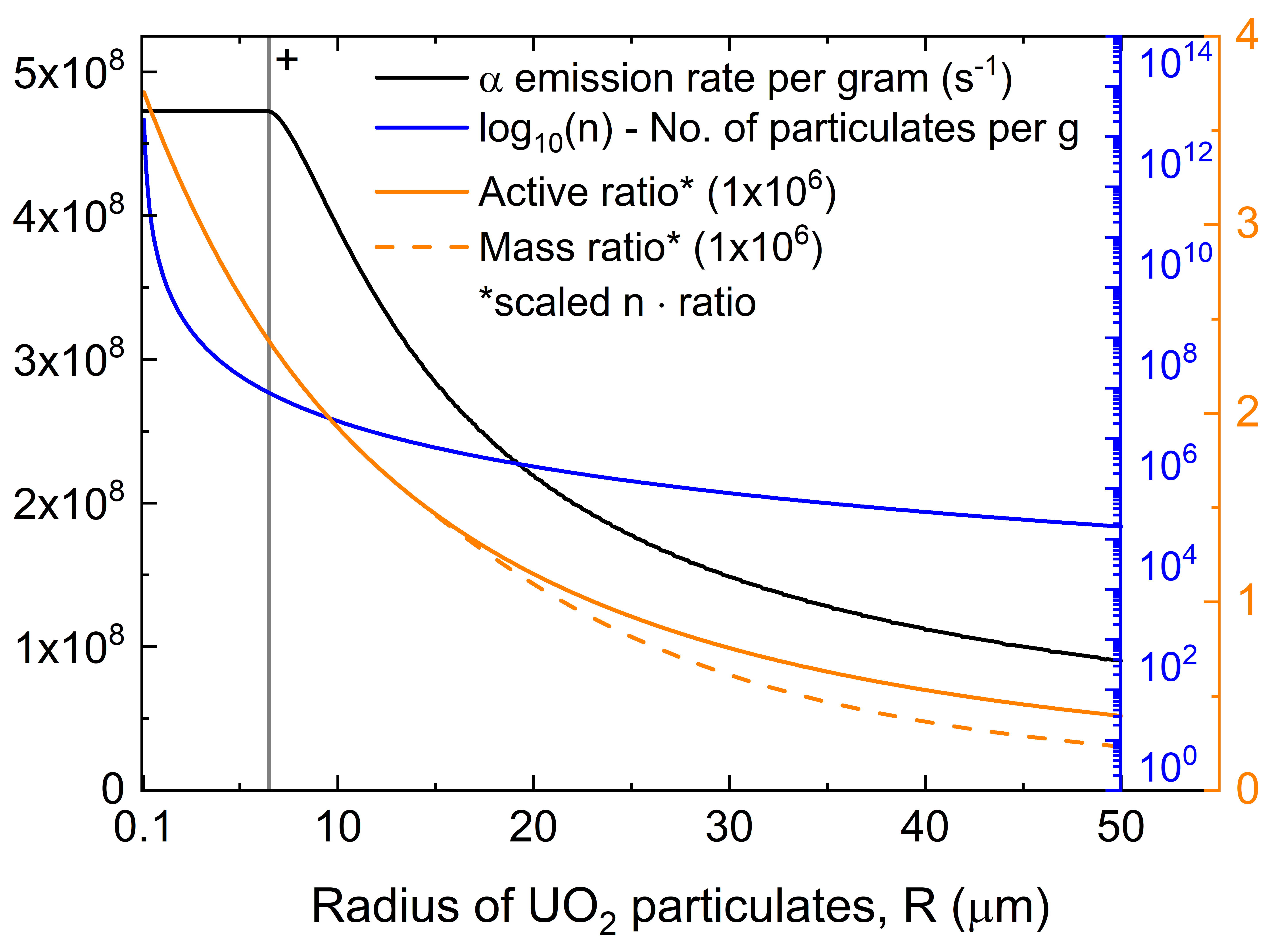}
 \caption{Graph illustrating the role of particulate sizes on dosimetry conditions when considering a unit (1 g) mass of UO$_2$ material in water. The black line indicates the rate of alpha particles emitted per gram of UO$_2$. The blue line indicates the number of particulates that make up 1 g of UO$_2$. The solid and dashed orange lines represent the active ratio and mass ratio, respectively, for 1 g of particulate mass.}
 \label{fig:1g}
\end{figure}
\indent The most significant motivation for this work is to be able to predict the radiolysis of water-suspended particulates of spent fuel of radii ranging from 1 $\mu$m to 50 $\mu$m. Where a perfectly dilute suspension of particulate fuel represents a worst-case scenario for radiolytic yields. When scaling the dose rate results to model more macroscopic scenarios of this type the particulate size in question is very important. Figure \ref{fig:perp} illustrates the relationship between the particulate size and  emission rate of alpha decays escaping the material. Included in this plot is the ratio of the `active mass’, the volume of material that if an $\alpha$-particle decays can escape, to `receiving mass’, that is the volume limit of water the flux can penetrate through surrounding the particulate, denoted `Active ratio’. The ratio of the total mass of the particulate to the receiving mass is plotted as `Mass ratio’. The mass ratio and  emission rate increase in a similar trajectory whereas the active ratio begins to fall in gradient beyond the 15-20 $\mu$m range. For isolated one particulate systems the influence of size is clear on the dose contributing factors. When applying the results to many particulate systems over a macro scale such as in the Sellafield fuel pond or K-basin uranic sludges a fixed mass approach is more suitable. Figure \ref{fig:1g} shows the effect of R on the active and mass ratio when normalised to 1 g of fuel material  including the scaling factor and emission rate. The scaled active and mass ratios represent a theoretical ratio of the total mass of material to surrounding water in a worst-case scenario of close packing particulates where all $\alpha$-decays travel through the fuel once. Essentially showing that although the active and mass ratios get increasingly smaller for smaller R, it is at a lower rate than the increasing number of particulates required if a fixed mass is considered.\\
\indent Considering Figure \ref{fig:perp} and Figure \ref{fig:1g} the reduced active and mass ratios for smaller particulate sized is reversed, illustrating a much greater efficiency in dose volumes considered. The  emission rate is also reversed but the peak  emission rate is bound by the activity of the material, only beginning to decay once the probability of alpha escape drops below 1, as expected. Note, the scaling factor will tend to infinity as R approaches 0, where, in order to maintain a perfectly dilute suspension of particulate fuel will tend to non-physical volumes considered; therefore, there is a limitation in considering these factors approaching the nanoscale. \\
\begin{figure}[!b]
\centering
\includegraphics[width=\linewidth]{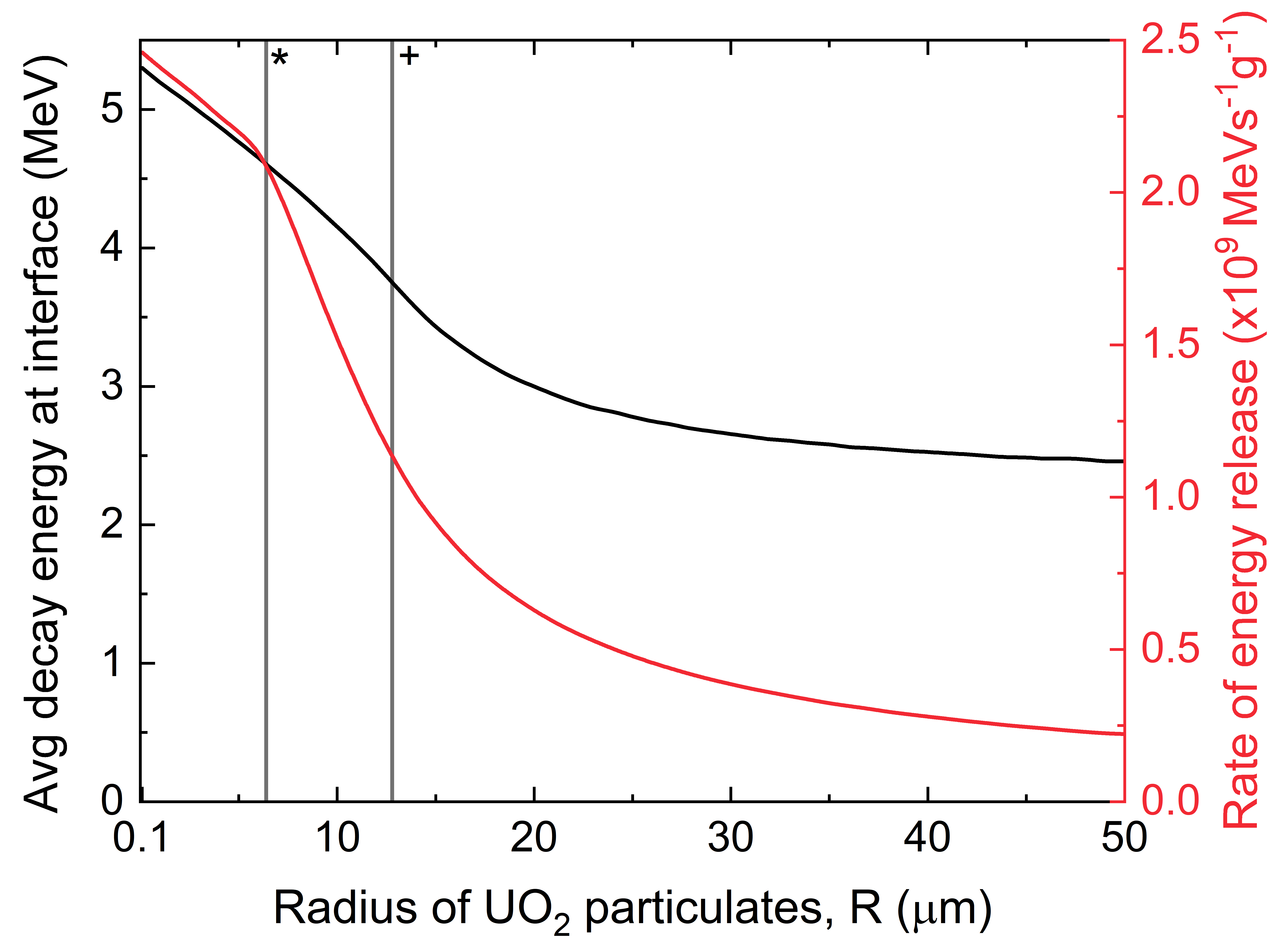}
\caption{Graph showing the role of increasing particulate size on the energy of $\alpha$-particles across the fuel-water interface and the rate of $\alpha$-decay energy released to the surrounding media from 1 g of UO$_2$. The activity and energy used in all studies were 4.73$\times 10^8$ Bqg$^{-1}$ and 5.3 MeV, respectively. The reference lines highlighted with a $^+$ and * indicates the limit where R$\leq \delta$ and R$\leq \delta/2$, respectively.}\label{fig:totenergy}
\end{figure}
\indent For macroscopic scenarios where radiolysis rates inside a large container are favoured over local concentrations around the fuel material, a more simplified approach is recommended. The results from this study can be distilled into two key results: 1) the average energy of an alpha decay at the interface, hence energy escaping the material, and 2) the total rate of energy escaping the fuel material, as shown in Figure \ref{fig:totenergy}. The total rate of energy release combines the interface energy results with the derived average probability of alpha escape. 

\section{Discussion}
The model presented in this study was tested against existing methods simulating $\alpha$-dosimetry in a spherical geometry. Previous attempts that model UO$_2$ particulates are presented by Tribet \textit{et al} (although the exact method was first presented by Mougnaud \textit{et al} considering nuclear glass) and Poulesquen \textit{et al} \cite{tribet2017spent,mougnaud2015glass,poulesquen2006spherical}. A direct comparison of results can be made between this study and the Tribet model \cite{tribet2017spent}. This is the only publication to use a radius below 100 $\mu$m through the use of the Monte Carlo n-particle code (MNCP). MCNP is built from FORTRAN, designed over 50 years ago but still commonly used for radiation modelling \cite{MCNP_hist}. Making the comparison of the results from this study and the MCNP code is important as the quality and robustness of MCNP has been proven through its wide use across the nuclear industry to this day. Figure \ref{fig:Tribet_comp} shows the dose rate profile for a 50 $\mu$m particulate from the Tribet study alongside this model. It can be concluded that the functional form and reduction in total dose rates are very similar between these models. The similarity in dose rate profile and the total dose rate lying within 0.7 \% of the Tribet result indicate comparable performance between the two. This suggests that the method first presented by Mougnaud \textit{et al.} on nuclear glass utilising MCNPX can produce accurate dosimetry results \cite{mougnaud2015glass}.\\
\indent For much larger particulates where a planar approximation becomes increasingly valid, comparisons were made for a 1 mm radius sphere in both the Poulesquen model and this study (Figure \ref{fig:large_comp}). The reduction of dose rate with water depth seen in the result by Poulesquen is much more severe than any model previously stated \cite{poulesquen2006spherical}. Poulesquen uses SRIM to calculate the projected range at 5.3 MeV and quotes it as $\delta_{UO_2}$ = 13.2 $\mu$m instead of the current 12.4 $\mu$m and $\delta_{H_2O}$ = 42 instead of 39.7 $\mu$m. This sensitivity of range data to dose rate results can be seen in a study comparing stopping power models by Hansson {et al.}, 2020 \cite{hansson2020alpha}. It shows that a few percent difference in range between ASTAR and SRIM stopping power data contributes to a negligible difference in the dose rate curve. This difference in range then fails to explain the scale of the increase and severity of decay shape in comparison to that of the Tribet model and the present study. This indicates an error in the dose calculation or an integral fault in the planar surface - spherical geometry combination. Poulesquen used a spherical geometry but takes the assumption $\delta$ << R. Leading to a semi-planar surface at a 1 mm sphere. This may be the reason for a dose drop-off that is much too sharp with an interface dose rate that is approximately twice that of the next highest interface dose rate reported in the literature \cite{poulesquen2006spherical}. The dose profiles shown in Figure \ref{fig:large_comp} illustrate the convergence to a linear solution in this study for large radii. The similarity between the large radii dose rate curve and the 8 mm diameter pellet modelled by Tribet \textit{et al.}, is clear. The slight discrepancies in dose rate shape between the two models may be due to a possible `edge effect' as mentioned by Tribet \cite{tribet2017spent}, due to the cylindrical shape of the pellet. \\
\indent The model was run incrementally to calculate the dose rate curve for particulates ranging from a radius of 1 $\mu$m to 50 $\mu$m. To understand how much energy is deposited as a function of spherical layers from the interface, the average LET per layer can be plotted and is shown in Figure \ref{fig:full_results}(a). Here it can be clearly seen that for smaller radii the Bragg curve of an $\alpha$-particle in water is less perturbed by the radial distance travelled in UO$_2$. It can be concluded that most of the energy is deposited away from the interface for particulates of R < 7 $\mu$m. This change in behaviour for different regimes of particulate size has not been previously reported in the literature.\\
\indent When using dosimetry results one must consider what is to be achieved from the following calculations. If for example, one wanted to consider dissolution kinetics of the radionuclide-containing interface, the commonly used dose rate curve is utilised. This becomes a more complex issue when particulates below 50 $\mu$m radius are used. This is because the volume differential becomes a significant feature in the dosimetry results (Figure \ref{fig:full_results}b). It is important to note that the hidden variable in the dose curves is the active ratio. The active ratio is the ratio of the mass of the material in the active layer to the mass of the receiving volume. Figure \ref{fig:perp} illustrates the change in gradient of the active ratio over the 1-50 $\mu$m range, plateauing for larger particulates. One could then consider the gap between the Active and Mass ratio (Figure \ref{fig:perp}) as the efficiency of decays escaping the active material, the smaller the gap the greater the efficiency. \\
\indent If one was to find the radiolytic production of a species in a solution containing particulates of radioactive material, such as in the Sellafield fuel pond sludges or K-basin sludges, a fixed mass approach is most suitable. In Figure \ref{fig:1g} the considerations from the isolated particulate are flipped. As expected the  emission rate remains constant for $R<\delta/ 2 $ and the curve follows the probability of alpha escape function shape beyond this.\\
\indent Although this presentation of data is useful for considering the macroscopic effects of geometry, the result begins to follow a nonphysical scenario for particulates towards the nanoscale. In this case, the number of particulates and volume of water considered for 1g would occupy a space too large to be applicable to the problems in question. \\
\indent To better understand the prevailing results from this study and the applicability for worst-case radiolytic production from alpha radiation in sludges, Figure \ref{fig:totenergy} is presented. The average energy upon crossing the fuel-water interface dictates energy escaping the fuel material.  Garisto (1989) derived an expression for the spectrum of energies emitted from a UO$_2$ planar surface as a function of decay energy, range and average rate of energy loss through the fuel \cite{GARISTO198933}. Using the derived average energy formula from this study

\begin{equation}
E_{avg} = \frac{A}{2R_0}\left[ \frac{2E_0^3}{3} + (\frac{B}{A})E_0^2 \right]
\end{equation}

where A = 0.358 $\mu$mMeV$^{-2}$, B = 1.2 $\mu$mMeV$^{-1}$, $E_0$ = 5.3 MeV and $R_0$ = 12.79 $\mu$m provides a result of $E_{avg}$ = 2.71 MeV. This result would be the asymptotic value for increasing particulate size if both studies were in agreement. Using a considerably larger radius than shown in Figure \ref{fig:totenergy} at R = 10 mm provides an average energy, E$_{avg}$ = 2.27 MeV. Suggesting the derivation by Garisto overestimates the average energy of the particles that emerge from a planar fuel surface.\\ 
\indent Combining the interface energy result with the derived probability of alpha escape function, the desired activity, and unit mass of material we arrive at the rate of alpha energy loss from the radionuclide emitting material, irrespective of surrounding media and ratio of masses (Figure \ref{fig:1g}). The average energy falls linearly from 5.3 MeV until the size of the particulate is half of $\delta$, then it increases in severity before plateauing again, similar to that of the P$_{\alpha}$ curve shape. This curve shape is then enhanced by the P$_{\alpha}$ function for the rate of energy release curve, resembling a decay curve beyond $\delta/2$, but falling at an almost constant rate below this.\\
\indent The results presented in this study indicate the importance of modelling particulates below 50 $\mu$m, as geometry has a pivotal role in the dose rates produced. The insights presented here enable a better understanding and prediction of chemical changes occurring in the local environment around particulates of spent nuclear fuel. 

\section{Conclusions}
A geometrical model, utilising analytical methods to build alpha dose rate profiles for small particulates of spent nuclear fuel has been presented. The novel approach uses a new derivation for the probability of alpha escape, emitted from a curved surface; considerably reducing the number of calculations required. The alpha dose rates were produced for particulates in the range of 1 $\mu$m to 10 mm in radius, showing considerable changes in the 1 $\mu$m to 50 $\mu$m range. Results also highlight issues with the presentation of this type of data, where careful consideration must be made over what result is required. For dosimetry and dissolution modelling the complex relationship between the mass of material, the `active' mass of material contributing to such a dose and the mass of water receiving such a dose must be considered in a spherical geometry. The mathematical method presented here could also be used for a multitude of radionuclide-containing materials and a variety of surrounding media as a multi-purpose tool for modelling radioactive materials.

\section*{Acknowledgements}
This work was supported by The Engineering and Physical Sciences Research Council (EPSRC) and the Transformative Science and Engineering for Nuclear Decommissioning (TRANSCEND) consortium.


\bibliographystyle{elsarticle-num}

\bibliography{Alpha_Spherical}

\end{document}